# Effects of the neutral point of dust charge in plasma sheath

**I. Driouch† * H. Chatei * M. El kaouini * M. El boujaddani * M. El hammouti ****

* *Department of Physics, Faculty of Science, University Mohammed I, BP 717, 60000 Oujda, Morocco,*
** *Department of Physics, Multidisciplinary Faculty, University Mohammed I, BP 300, 62700 Nador, Morocco*

†* Corresponding author.
E-mail: *driouch_ismael@yahoo.fr*

**Abstract:** We numerically investigate the dust charging in the sheath, by using the usual fluid approximation, it is extended to include self consistently the dust charge variation. The grain charge becomes a new self consistent dynamic variable, it is found that dust grains are first charged negatively at the sheath edge and then begin to be charged positively in the sheath. The numerical results show that the initial velocity of dust charged grains, sized grains and the density ratio of ion to electron at the sheath edge have affected the neutral point of dust charge. Moreover, the neutral point of dust charge affects the spatial distribution of dust density.

*Keywords:* fluid approximation, sheath modeling, dusty plasma sheath.

## 1. INTRODUCTION

The study of the dusty plasma sheath is one of the important problems in the experiment equipments of plasma. So the correlative works both of experiment and theory are developed widely during the past several years [1-10]. Recently, the sheath of dusty plasma has been studied with constant charge dust grains [3,4]. Many authors have investigated the problems to include the effects of dust charge variation [1, 9-12], the charges of dust grains are assumed to be determined only by a balance of the electron and ion currents flowing into the grains. Wang Zheng-Xiong et al. studied the effects of ion temperature [14] and the negative ions on the zero-point of dust charging (neutral point of the dust charge) [15] in the plasma sheath. In Ref.[9], Jin.Yuan.Liu et al. studied the distribution and suspension of the particles in a plasma sheath by the fluid model and self-consistent grain charge variations. In their study, only the negative charged particles were taken into consideration.

In this study, we report a numerical method for the solution of the sheath modification created in the presence of cold fluid dust grains with variable charge. In addition, we study the dust motion and suspension in the sheath by taking into consideration the neutral point of dust charge. We shall assume that the charge variation of the dust is determined self consistently as in the Ref.[16]. For the conditions when the mean free path is long and the dust radius is small compared to the Debye length, the orbit motion limited theory [17,18] will be used to calculate the charging currents. A fluid model and self-consistent grain charge variations are chosen to describe the dusty plasma sheath, composed of electrons, ions and charged dust grains with radius in order of nanometers size. We have considered the effect of parameters which modify the dusty plasma sheath characteristics, such as the action of dust grain radius, the density ratio of ions to electrons and the initial velocity of dust. These fundamental parameters affect significantly the zero point of the dust charge, this neutral point affects effectively the spatial dust density distribution. This work is organized in the flowing fashion. In sec.2 The analytical model of the sheath structure and the dust charging in the sheath are presented. In sec. 3 the numerical results are presented and discussed. Finally, a conclusion is given in sec 4.

## 2. DUSTY PLASMA SHEATH MODELS

We consider a stationary state ($\partial_t = 0$), collisionless and unmagnetized dusty plasma sheath in contact with a planar wall, consisting of electrons, ions and nano size charged grains, which has one-dimensional coordinate space. Assume the sheath region lies between $x = 0$ (the sheath edge), and the wall can be located any where in the region $x > 0$. At the sheath edge $x = 0$, the potential is assumed to be taken zero, the electron, ion and dust densities $n_{e0}, n_{i0}, n_{d0}$ satisfy the quasi-neutrality condition $e(n_{i0} - n_{e0}) - q_{d0}n_{d0} = 0$, here $e$ is the electron charge and $q_{d0}$ the dust charge at the sheath edge. The electrons are assumed to be in thermal equilibrium state, thus the density $n_e$ is given by the Maxwell-Boltzmann distribution.

$$n_e = n_{e0} \exp(\frac{e\phi}{T_e}) \quad (1)$$

where $\phi$ is the spatial electric potential in the sheath, $T_e$ is the electron temperature and $n_{e0}$ is the electron density at the sheath edge.

The ions, are treated as a cold fluid, governed by the continuity and the momentum conservation equations.

$$u_i n_i = u_{i0} n_{i0} \quad (2)$$

$$m_i u_i \frac{\partial u_i}{\partial x} = -e\frac{\partial \phi}{\partial x} \quad (3)$$

where $u_i$, $n_i$ and $m_i$ are velocity, density and mass of the ions in the sheath, respectively.

The dust grains are also treated as cold fluid obeys the sources free, the continuity and momentum equations are.

$$u_d n_d = u_{d0} n_{d0} \quad (4)$$

$$m_d u_d \frac{\partial u_d}{\partial x} = -q_d\frac{\partial \phi}{\partial x} \quad (5)$$

where $q_d = -ez_d$ is the variable-charge of dust grains, $z_d$ is the charge number of the dust, $m_d$, $n_d$ and $u_d$ are respectively the dust particle mass, density and velocity of the dust grains. The dust charge arises from plasma currents due to the electrons and the ions reaching the dust grains surface. In this case, the variable dust charged is determined self-consistently by the charge conservation.

$$u_d \frac{\partial q_d}{\partial x} = I_e + I_i \quad (6)$$

where $I_e$ and $I_i$ are the plasma electron and ion currents, respectively, flowing the dust grains surface. The charging currents are from the thermal electrons and the cold ions [19,20].

$$I_e = -\pi r_d^2 e n_e (\frac{8T_e}{\pi m_e})^{\frac{1}{2}} \exp(\frac{e\phi_d}{T_e}) \quad (7)$$

$$I_i = \pi r_d^2 e u_i n_i (1 - \frac{2e\phi_d}{m_i u_i^2}) \quad (8)$$

where $r_d$ is the dust grain radius, $m_e$ is the electron mass, $\phi_d$ the dust surface potential. Then, the dust charge $q_d$ can be determined for a given $r_d$ by the dust charge surface potential relation $\phi_d = \frac{q_d}{r_d}$.

The system of equations is completed by the Poisson's equation.

$$\frac{\partial^2 \phi}{\partial x^2} = -4\pi(e(n_i - n_e) + q_d n_d) \quad (9)$$

In order to solve the system equations $(1)-(9)$, we introduce new dimensionless variables $\eta = -\frac{e\phi}{T_e}$ is the normalized sheath potential, the normalized dust charge $\eta_d = \frac{-q_d}{ze}$, the density ratio of ion to electron $\delta_{ie} = \frac{n_{i0}}{n_{e0}}$ and the density ratio $\delta_{de} = \frac{n_{d0}}{n_{e0}}$, as well as normalize the coordinate by $\xi = \frac{x}{\lambda_D}$ with respect to the electron Debye length $\lambda_D = (\frac{T_e}{4\pi e^2 n_{e0}})^{1/2}$. We also normalized ion and dust velocities $v_d = \frac{u_d}{c_d}$ , $v_i = \frac{u_i}{c_i}$, respectively.
Here $c_d = (\frac{zT_e}{m_d})$ is the dust-acoustic speed, with $z = \frac{r_d T_e}{e^2}$, $c_i = (\frac{T_e}{m_i})$ is the ion-acoustic speed. Substituting these dimensionless variables into equations $(1)-(9)$, we obtain

$$N_e = exp(-\eta) \quad (10)$$

$$\frac{\partial N_i}{\partial \xi} = -\frac{N_i}{v_i^2}\frac{\partial \eta}{\partial \xi} \quad (11)$$

$$N_d = \frac{v_{d0}}{v_d} \quad (12)$$

$$\frac{\partial v_i}{\partial \xi} = \frac{1}{v_i}\frac{\partial \eta}{\partial \xi} \quad (13)$$

$$\frac{\partial v_d}{\partial \xi} = -\frac{\eta_d}{v_d}\frac{\partial \eta}{\partial \xi} \quad (14)$$

$$\frac{\partial \eta_d}{\partial \xi} = \frac{\Re}{v_i}[\delta_{ei}\beta \exp(-\eta - \eta_d) - N_i v_i (1 + \frac{2\eta_d}{v_i^2})] \quad (15)$$

$$\frac{\partial^2 \eta}{\partial \xi^2} = \delta_{ie} N_i - \exp(-\eta) - \frac{v_{d0}}{v_d} z \eta_d \delta_{de} \quad (16)$$

where $\Re = \frac{\mu^{1/2}\pi r_d^2 \lambda_D n_{i0}}{z^{3/2}}$, $\mu = \frac{m_d}{m_i}$ and $\beta = (\frac{8m_i}{\pi m_e})^{1/2}$.

## 3. NUMERICAL RESULTS

We have numerically solved equations $(10)-(16)$. In the flowing calculations, we take some typical plasma parameters that are representative of argon plasma, such as $n_{e0} = 3 * 10^{10}(cm^{-3})$, $T_e = 2(eV)$. The starting point is taken at $\xi = 0$, where we set $\eta = 0$, $\frac{\partial \eta}{\partial \xi} = 0.01$, $v_{i0} = 1.1$. We consider the dust spherical particles of radius is in the range of $50 - 200nm$ with uniform mass density $\rho_d = 2.2(g/cm^3)$ so that $T_e/e^2 = T_e/1.4 * 10^{-9} \approx 1.43 * 10^9(m)$ for $T_e = 2(eV)$.
The results are presented in the following figures.

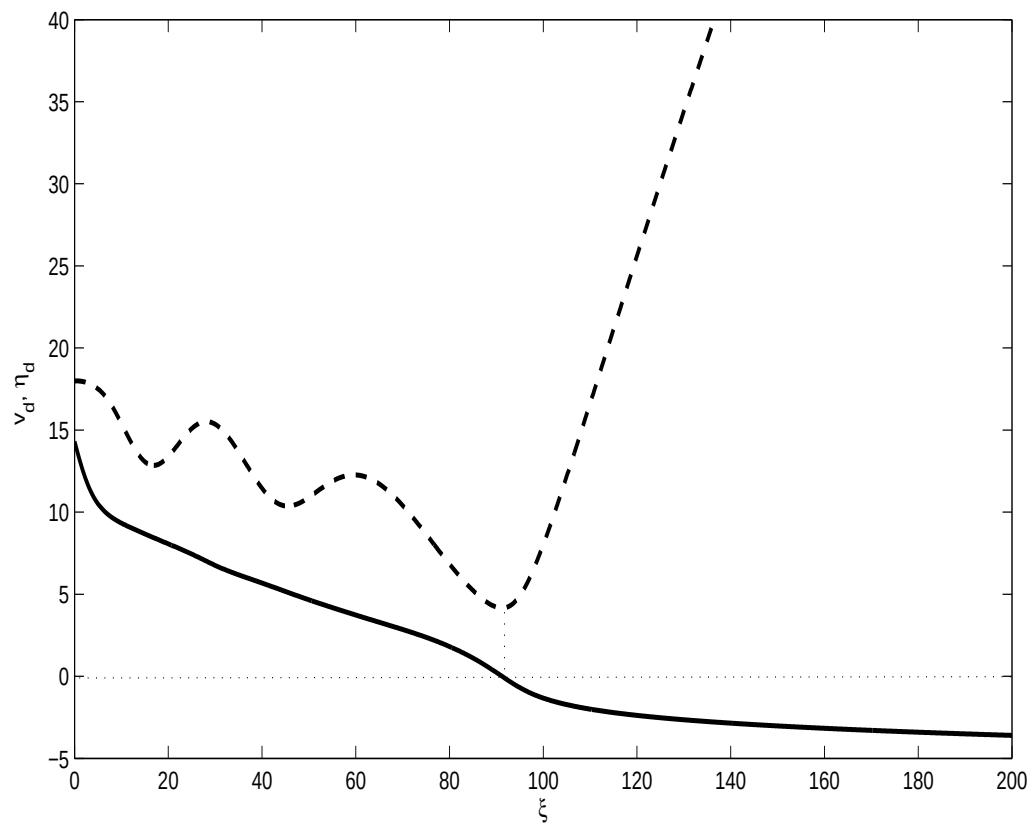

Fig. 1. The normalized distributions of dust charge (solid line) and dust velocity (dashed line) versus $\xi$ (The normalized position) in the plasma sheath for parameters $\delta_{ie} = 1.8$, $v_{d0} = 18$ and $v_{i0} = 1.1$.

Figure 1 shows the normalized dust charge and dust velocity distributions versus $\xi$ (normalized position) in the sheath for $r_d = 100nm$, $\delta_{ie} = 1.8$ and $v_{d0} = 18$. It is shown that the dust charge is negative at the sheath edge, and when the dust particles are moving to the wall, the dust charge decreases to zero, then close to the wall they are charged positively. This is due to the fact that the density of electrons falls faster than that of ions in the sheath.

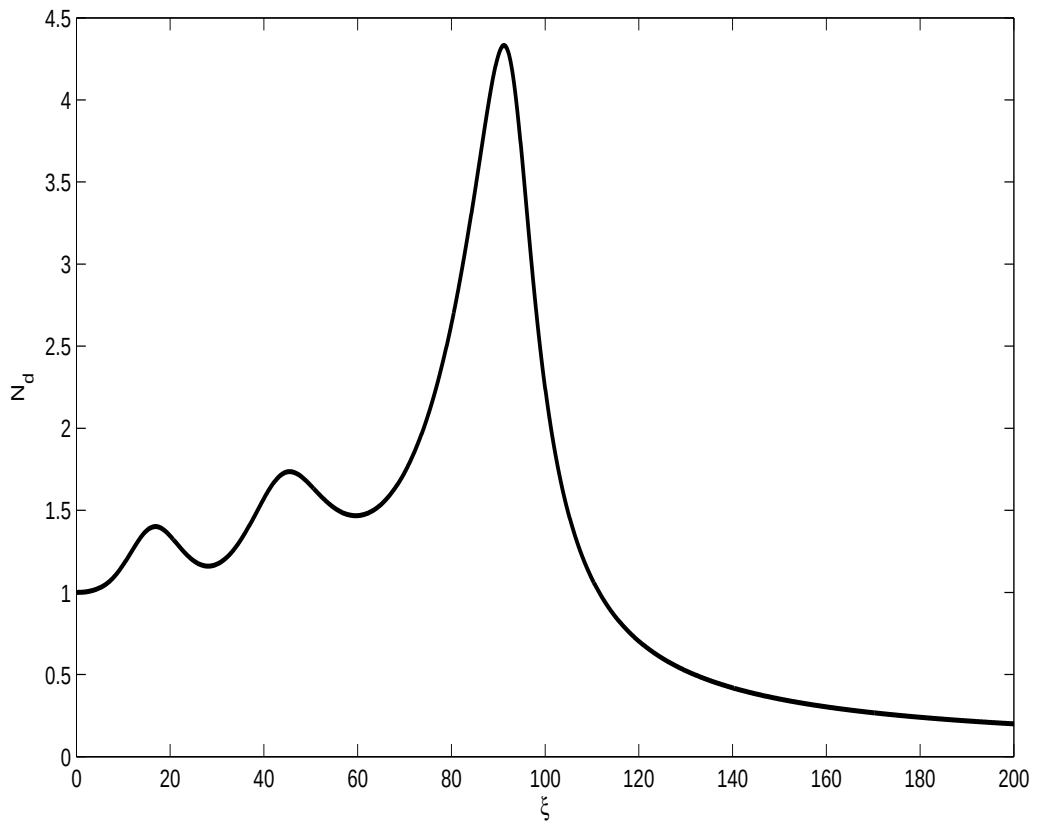

Fig. 2. The normalized dust density distribution versus $\xi$ (normalized position) in the plasma sheath with the same parameters as figure 1.

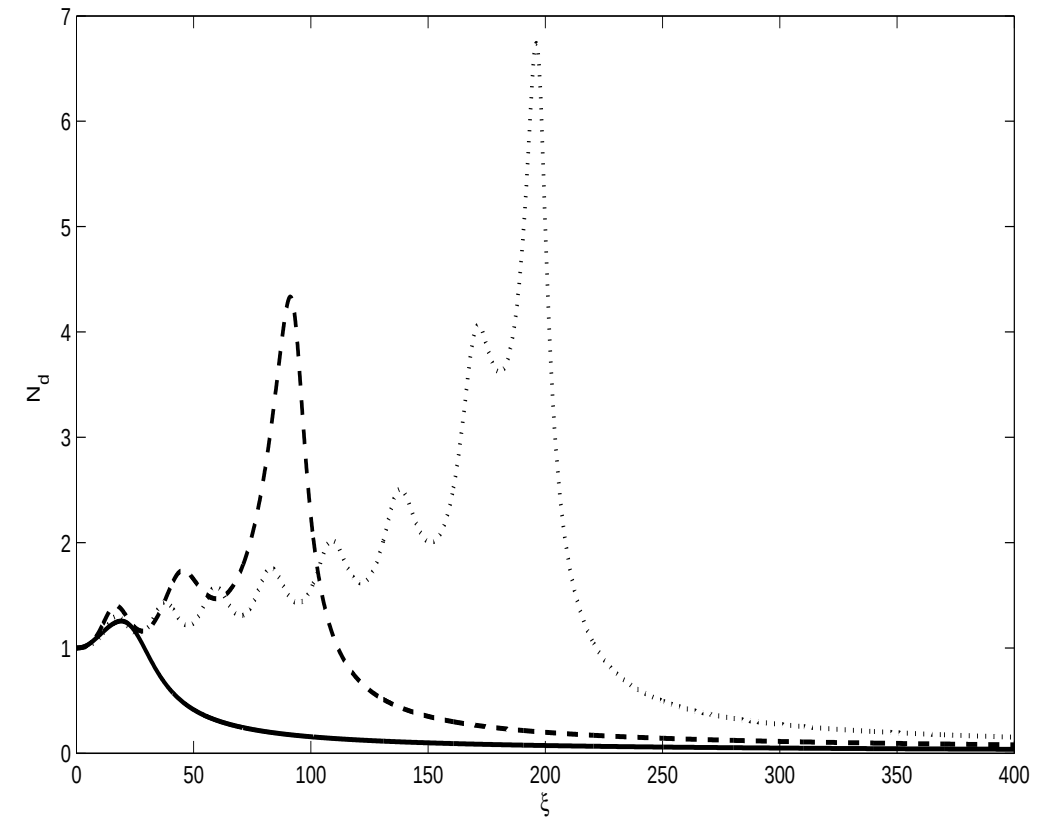

Fig. 3. The normalized dust density distribution versus $\xi$ (normalized position) in the plasma sheath with the same parameters as figure 1 and different values of dust grains radius $r_d = 200nm$ (solid line), $r_d = 100nm$ (dashed line) , $r_d = 70nm$ (dotted line).

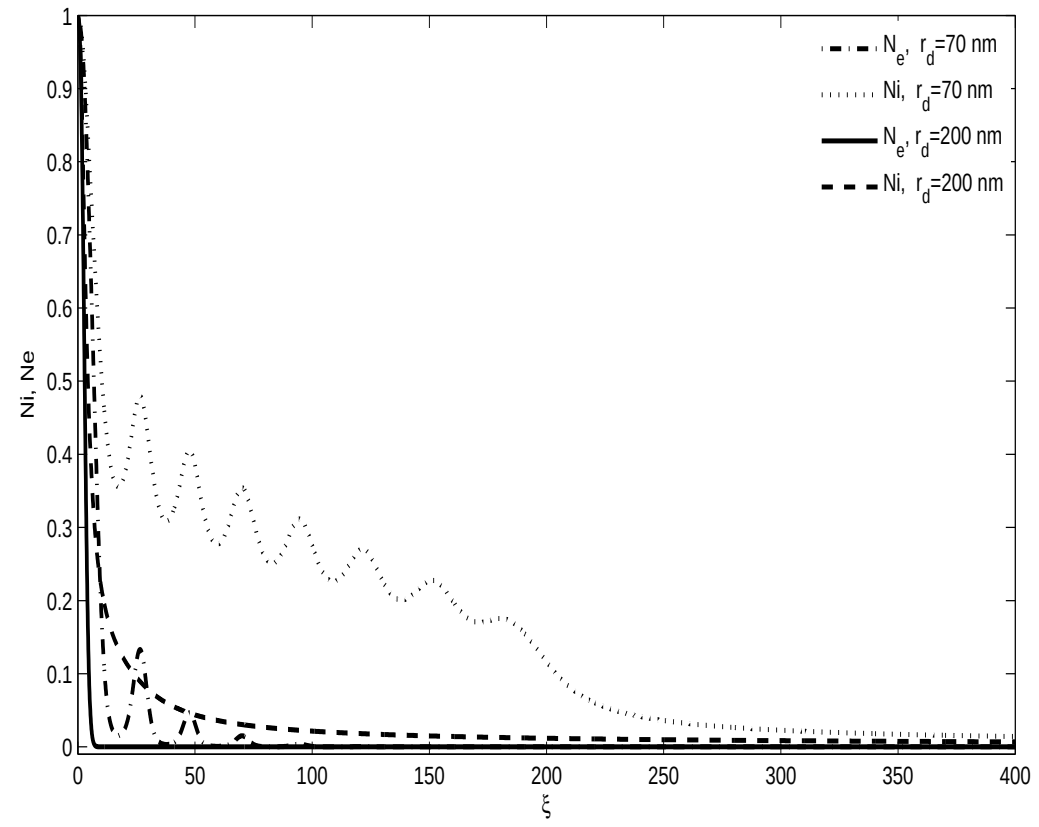

Fig. 4. The normalized density distributions of ions and electrons versus $\xi$ (normalized position) in the plasma sheath with different values of dust grains radius $r_d = 200nm$ $r_d = 70nm$.

Then, the grain charge is due only to the ion flux, where the electrons vanish and there is a pure ion sheath near the wall. The curve of the dust charge reveals the existence of a neutral point and the dust velocity curve reveals the existence of a minimum at this position, this is because the dust velocity and dust charge are related in the sheath by the momentum equation of dust( Eq.14). It is clear, from this equation, that the dust velocity represents an extremum at neutral point of dust charge $\eta_d = 0$.

Figure 2 shows the distribution of dust density in the sheath for the conditions same as for fig.1. It takes on the oscillatory structures at first, then sharp peak emerges and rapidly drops. The result is the same as J.Y.Liu's result[8]. The sharp peak emerges at zero point of dust charge where the dust velocity is minimum.

Figure 3 shows the dust density distribution under various values of the dust grain radius $r_d$, while the equilibrium dust velocity is $v_{d0} = 18$. We notice that, the dust density distribution shows a peak, which indicates that in this region, more dust particles are gathered. With the decrease of dust radius, the peak of the dust density moves towards to the wall. This is because the zero point of the dust grain charge for large dust grain radius closes faster to the sheath edge in opposite to small grain radius. Thus, the region of dust disperse becomes wider.
Figure 4 shows comparisons between density distributions of electrons and ions for two cases $r_d = 70nm$ and $r_d = 200nm$ in the sheath. Both the ion density $N_i$ and the electron density $N_e$ fall slowly in the case of ($r_d = 70nm$) compared to the case of ($r_d = 200nm$). For two cases from a certain position the electrons begin to vanish and the sheath become an ions.

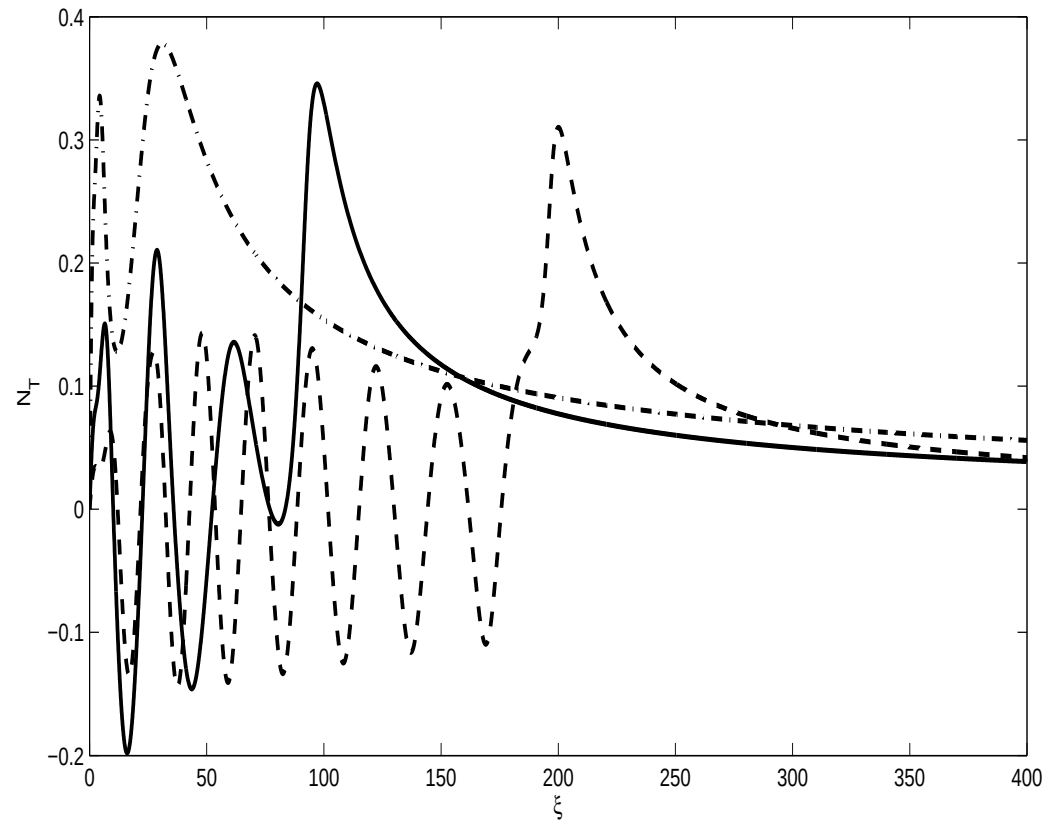

Fig. 5. The normalized net space charge density versus $\xi$ (normalized position) in the plasma sheath with the same parameters as figure 1 and different values of dust grains radius $r_d = 200nm$ (solid line), $r_d = 100nm$ (dashed dot line), $r_d = 70nm$ (dashed line).

Figure 5 shows the net space charge for various values of the dust grain radius $r_d$. The net space charge takes the oscillatory structures at first, then the positively charged gather in certain region, so the net space charge density in such region of the sheath emerges at a peak value. The case of the dust radius ($r_d = 200nm$) shows the net space charge increases fast at the sheath edge, because the electron and negatively charged dust vanish more rapidly than those in the cases of radii ($r_d = 100nm$) and ($r_d = 70nm$). Thus, the net space charge is greater with increasing the radius of dust. Accordingly the spatial potential in the sheath increases more rapidly for dust of large radii (see fig.6), because of its high net space charge density.

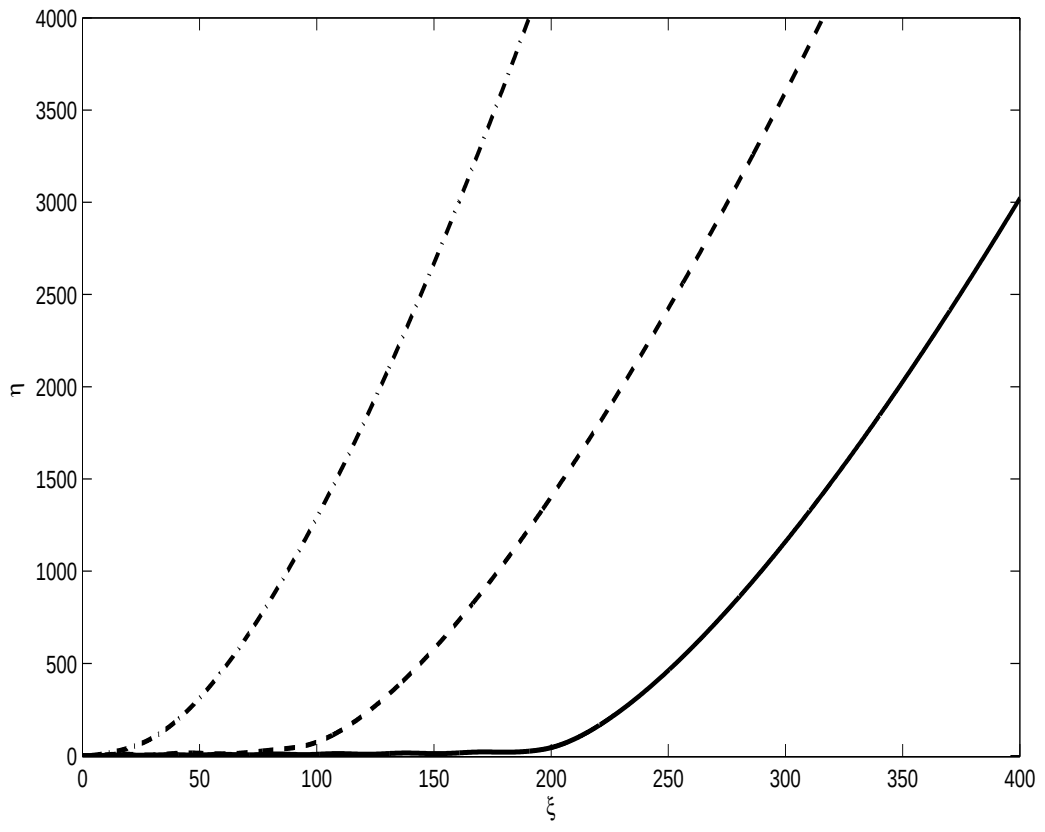

Fig. 6. The normalized spatial electric potential versus $\xi$ (normalized position) in the plasma sheath with the same parameters as figure 1 and different values of dust grains radius $r_d = 70nm$ (solid line), $r_d = 100nm$ (dashed line),$r_d = 200nm$ (dashed dot line).

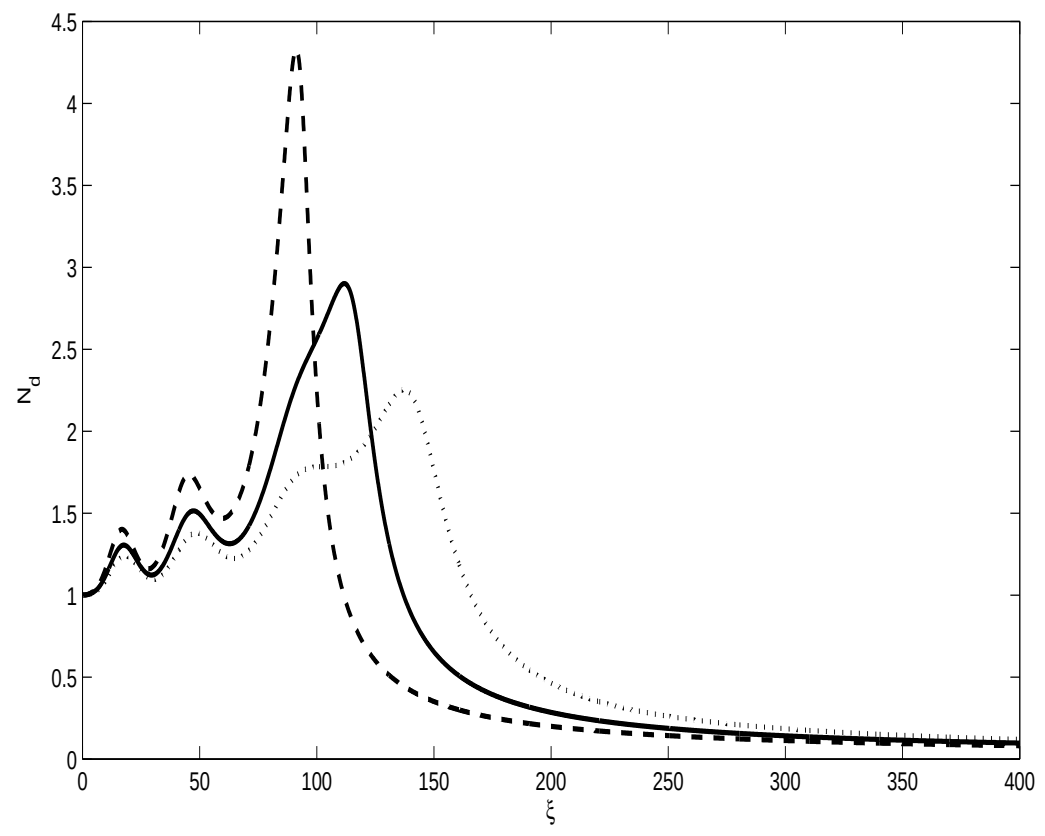

Fig. 7. The normalized dust density distribution versus $\xi$ (normalized distance) in the plasma sheath with the same parameters as figure 1 for various dust velocity at the sheath edge $v_{d0} = 20$ (solid line), $v_{d0} = 18$ (dashed line), $v_{d0} = 22$ (dotted line).

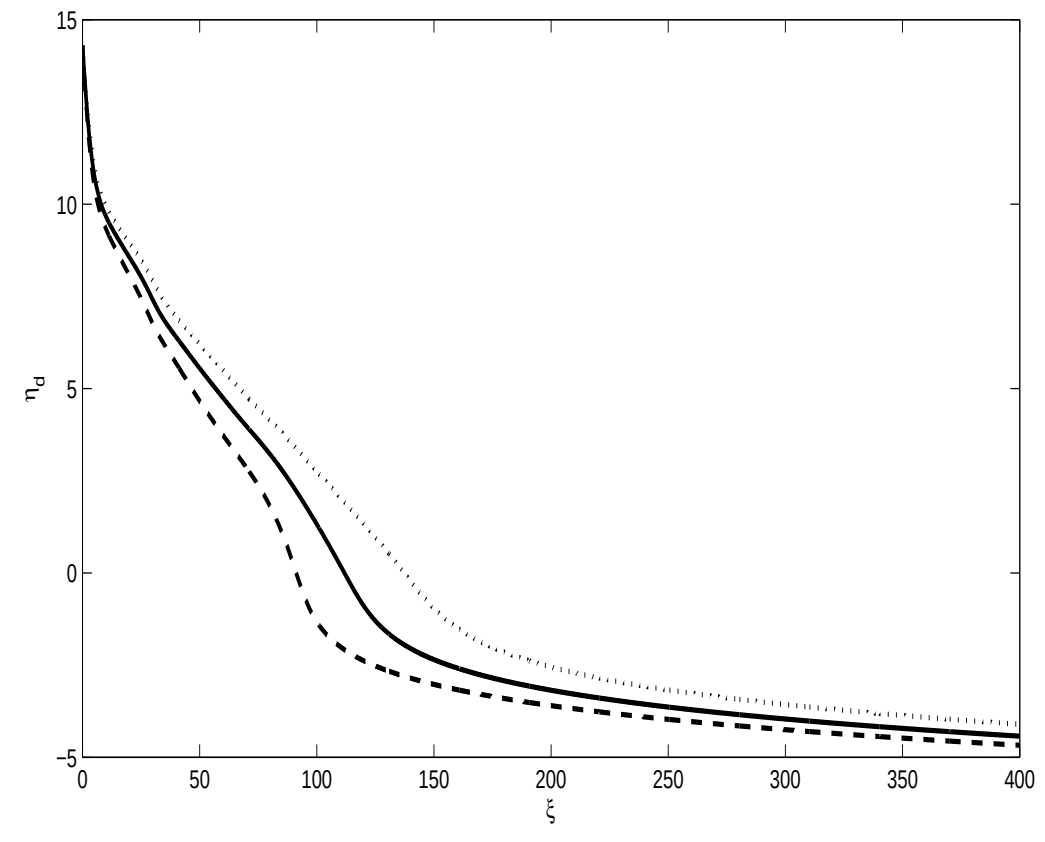

Fig. 8. The normalized dust charge distribution versus $\xi$ (normalized distance) in the plasma sheath with the same parameters as figure 1 for various dust velocity at the sheath edge $v_{d0} = 20$ (solid line), $v_{d0} = 18$ (dashed line), $v_{d0} = 22$ (dotted line).

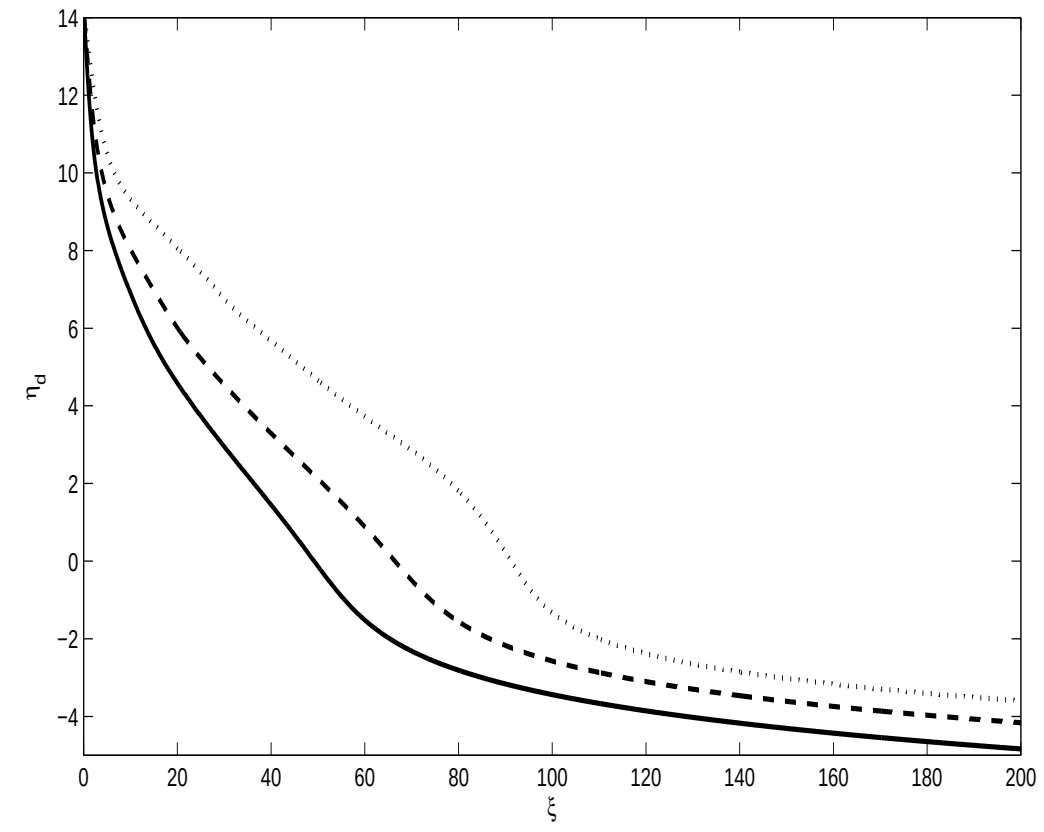

Fig. 9. The normalized dust charge distribution versus $\xi$ (normalized distance) in the plasma sheath with the same parameters as figure 1 for various density ratio of ion to electron at the sheath edge $\delta_{ie} = 3.8$ (solid line), $\delta_{ie} = 2.8$ (dashed line), $\delta_{ie} = 1.8$ (dotted line).

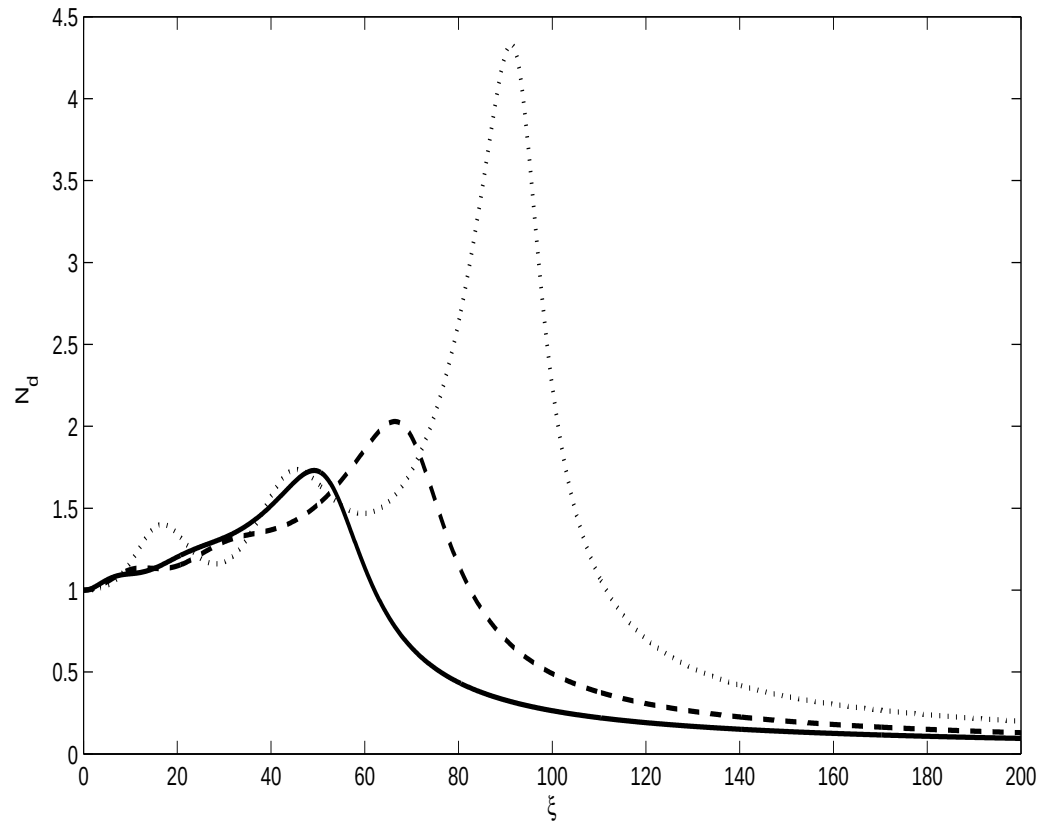

Fig. 10. The normalized dust density distribution versus $\xi$ (normalized distance) in the plasma sheath with the same parameters as figure 1 and for various density ratio of ion to electron at the sheath edge $\delta_{ie} = 3.8$ (solid line), $\delta_{ie} = 2.8$ (dashed line), $\delta_{ie} = 1.8$ (dotted line).

In figures 7-8, we study the effect of the parameter $v_{d0}$ (the equilibrium dust velocity) on dust charge and density distributions. It is shown that the bigger of the dust density nearer from the sheath edge. It is also found that greater values of $v_{d0}$ favor the expansion of the region of dust disperse, this make distribution of dust approach to the wall fig.7. This is because the neutral point of dust charged in the case of $v_{d0} = 18$ is closer to the sheath edge than that in the cases of $v_{d0} = 20$ and $v_{d0} = 22$ , the reason for it is that the negatively charged particles of $v_{d0} = 18$ vanish faster than that in the cases of $v_{d0} = 20$ and $v_{d0} = 22$ see fig.8.

The effect of the density ratio of ion to electron (the equilibrium dust density) on the dust density and the dust charge distributions in the sheath is studied in the figures 9-10. By increasing value of $\delta_{ie}$, there are more ions than electrons at the sheath edge, so the dust grains are less negative and for that the grains charge drops quickly to zero. One can see that by increasing the density ratio of ion to electron $\delta_{ie}$, i.e much ions at the sheath edge. It is shown that for the great value of $\delta_{ie} = 3.8$ compared to $\delta_{ie} = 1.8$, the neutral point of dust charge closes faster near the sheath edge. Accordingly the position of the peak of dust density shifts towards the wall for the small values of $\delta_{ie}$ fig.9. One can see also that the peak of dust density decreases with increasing $\delta_{ie}$, this can be explained by the fact that at the sheath edge, the dust density $n_{d0}$ is given by the quasi neutrality condition $(n_{d0} = (\delta_{ie} - 1)/z_{d0})$, so that the normalized density $N_d(= n_d/n_{d0})$ is low.

## 4. CONCLUSION

In this paper, we have presented the spatial distribution and suspension of the nanodust particles with self-consistent charge variation in plasma sheath using fluid model. It is found that the presence of dust can lead to a very different physical behavior compared to that of the electron-ion plasma. The details of the dust particles, such as the variations in size, charge and mass can affect the spatial distribution of dust. Furthermore, the dust grains with different initial velocities, sizes and densities entering the sheath, show different motion states: returning at the sheath edge and moving at the wall, thus; the region of dust disperse is modified. In addition, it is shown that at the sheath edge the dust particles are charged negatively then, it are become charged positively in the sheath. Moreover, we also found that the neutral point of dust charge shifts by different boundary conditions at the sheath edge. The results of the present investigation can be used in understanding that the neutral point of the dust charge play an important role on the dust particles motion in the sheath.

## REFERENCES

[1] Z. X. Wang, X. Wang, J. Y. Liu and Y. Liu, *J. Apll. Phys* 97 (2005) 1–5.
[2] J. X. Ma and M. Y. Yu, *Phys. Plasmas* 2 (1995) 1343-1345.
[3] M. Y. Yu, H. Saleema and H. Luo,*Phys. Fluids B* 4 (1992) 3427-3431.
[4] M. K. Mahanta, K. S. Goswami,*Pramana-J. Phys* 56 (2001) 579-584.
[5] D. Tskhakaya , P. K. Shukla , B. Eliasson, *Physics Letters A* 331 (2004) 404–408.
[6] T. E. Sheridan,*J. Appl. Phys* 98 (2005) 2-6.
[7] T. Nitter, *Plasma Source Sci. Technol* 5 (1996) 93–111.
[8] C. Arnas, M. Mikikian, G. Bachet, and F. Doveil, J. X. Ma and C. X. Yu, *Phys. Plasmas* 7 (2000) 4418-4422.
[9] J. Y. Lieu, D. Wang and T. C. Ma, *Vacuum* 59 (2000) 126–134.
[10] D. Wang and M. Sheng, *J. Appl. Phys* 94 (2003) 1368-1373.
[11] D. Wang and X. Wang,*J. Appl. Phys* 89 (2001) 3602-3605.
[12] N.-C. Wang, B.-S. Xie, *Phys. Plasmas* 9 (2002) 717-720.
[13] C. Lin, Wei-gui Feng, M. M. Lin, *Communications in Nonlinear Science and Numerical Simulation* 13 (2008) 1287–1293.
[14] Z. X. Wang, J. Liu, Y. Liu and X. Wang, *Chin. Phys. Lett*, 21 (2004) 697-699.
[15] Wang Zheng-Xiong, Wang Wen -Chun, Liu Yue, Liu Jin-Yuan and Wang Xiao-Gang, *J. Plasma. Physics* 70 (2004) 577-581.
[16] M. Tribeche, H. Houili, and T. H. Zerguini, *Phys. Plasmas* 9 (2002) 419-429.
[17] E.C. Whipple, *Rep. Prog. Phys* 44 (1981) 1198–1250.
[18] J. E. Allen, *Physica Scripta* 45 (1992) 497–503.
[19] A. Barkan, N. D'Angelo and R. L. Merlino, *Physical Review Letters* 73 (1994) 3093–3096.
[20] J. Goree, *Plasma Source Sci. Technol* 3 (1994) 400–406.